\documentclass[journal]{IEEEtran}
\usepackage{cite}
\usepackage{amsmath,amssymb,amsfonts}
\usepackage{algorithmicx}
\usepackage{algorithm}
\usepackage{graphicx}
\usepackage{textcomp}
\usepackage{xcolor}
\usepackage{amsmath} 
\usepackage{bm}
\usepackage{float}
\usepackage{subcaption}
\usepackage{tikz}
\usepackage{pgflibraryarrows}
\usepackage{pgflibrarysnakes}
\usepackage{xcolor}
\usepackage{tikz-3dplot}
\usepackage[export]{adjustbox}
\usepackage{pgfplots}
\usetikzlibrary{positioning}
\definecolor{meta_col}{rgb}{0, 0, 0}
\definecolor{irs_col}{rgb}{0, 1, 0}

\DeclareMathOperator{\acot}{acot}
\definecolor{v_col}{rgb}{0,0.4470,0.7410} 

\def\BibTeX{{\rm B\kern-.05em{\sc i\kern-.025em b}\kern-.08em
		T\kern-.1667em\lower.7ex\hbox{E}\kern-.125emX}}
\makeatletter
\tikzset{
	dot diameter/.store in=\dot@diameter,
	dot diameter=1pt,
	dot spacing/.store in=\dot@spacing,
	dot spacing=0.5pt,
	dots/.style={
		line width=\dot@diameter,
		line cap=round,
		dash pattern=on 0pt off \dot@spacing

	}
}

\begin{document}
\title{Differential Polarization Shift Keying Through Reconfigurable Intelligent Surfaces}	
	
\author{Emad~Ibrahim,
	Rickard~Nilsson, and~Jaap~van~de~Beek
	\thanks{The authors are with the Department of Computer Science, Electrical
		and Space Engineering, Luleå University of Technology, 97187 Luleå,
		Sweden e-mail:\{emad.farouk.ibrahim, rickard.o.nilsson, jaap.vandebeek\}@ltu.se}
	}
\maketitle
\begin{abstract}
We propose a novel reconfigurable intelligent surface (RIS) encoded information transmission scheme for a line-of-sight environment. A RIS fed with data modulates the information on impinging waves emitted from an external source in the states of polarization (SoP) of the scattered waves by performing a novel differential polarization shift keying. In particular, the information is encoded in the change of the SoP over two successive scattering slots. The proposed scheme is immune to the SoP fluctuations in the wireless channel which allows for non-coherent detection at the receiver.


\end{abstract}
\section{Introduction}
\let\thefootnote\relax\footnote{We acknowledge the partial funding by the EU's Interreg Nord Program.}
Reconfigurable intelligent surface (RIS) has been introduced as attractive energy-efficient hardware for wireless communications applications. A RIS is a thin planar surface of multiple reflecting units each of which has a tunable interaction with the incident waves. Conventionally, the RIS units induce independently controllable reflection coefficients to the scattered waves on them. Therefore, the RIS is capable of re-engineering continuously the wireless channel between the transmitter and receiver \cite{b1}. 

One of the promising use cases of the RIS is to act as an access point for information transmission wherein the RIS relies on an external radio frequency (RF) source to modulate the data that arises from a source attached to it. Specifically, in \cite{b2}, the on/off states of the RIS units are alternated to perform a sort of amplitude modulation which comes at expense of the RIS beamforming capability as some units become inactive. Furthermore, \cite{b3} proposed RIS encoded index modulation which is limited to rich-scattering environments wherein low channel correlation among different antenna indices is guaranteed. However, it becomes infeasible in the line-of-sight (LoS) environment of highly correlated channels. Moreover, \cite{b4,b5} proposed RIS encoded reflecting modulation wherein the data are encoded in the reflection patterns of the RIS. However, the selected reflection patterns cannot simultaneously achieve full beamforming gains for different modulated symbols given the wireless channel variations and because some RIS units are turned off therein.

On the other hand, one of the attractive functions of the RIS is that it can control the state of polarization (SoP) of the scattered waves from its units \cite{b6}. The SoP of the electromagnetic wave is defined by the orientation of the propagating wave relative to the direction of propagation \cite{b7}. Since any SoP can be decomposed into any arbitrary orthogonal SoPs pair. Accordingly, a RIS of dual-polarized (DP) units, which excites orthogonal SoPs pair and induces independent phase shift per each SoP, is capable of tuning the SoP of scattered waves \cite{b8,b9}.

In this letter, we propose a novel RIS encoded information transmission scheme for a LoS environment by relying on the SoP tuning capability of the RIS. In our earlier work \cite{b10}, we developed a conventional polarization shift keying (CPolSK) modulation scheme wherein the information is encoded directly in the SoP of the scattered wave from the RIS. However, in order not to sacrifice performance, the receiver needs to properly estimate and compensate for the SoP fluctuations that occur in the wireless channel, which becomes mainly an SoP rotation given the LoS environment. To solve this issue we develop a differential PolSK (DPolSK) modulation scheme. The RIS is first tuned to achieve full beamforming towards the receiver to overcome the large path losses to and from the RIS in a far-field scenario \cite{b11}. The information is then encoded in the change of the SoP over two successive scattering slots, by either preserving or switching the SoP of the current slot in comparison to that in the previous slot. For data detection, the receiver tracks the SoPs over two successive slots. Given that the wireless channel is static over two consecutive scattering slots, the proposed scheme becomes immune to the SoP fluctuations in the wireless channel. It enables non-coherent detection without the need for SoP rotation estimation and compensation at the receiver.

In \cite{b12,b13} the authors proposed a RIS assisted differential reflecting modulation scheme wherein the information is encoded in the permutation order of the RIS reflection patterns. However, the chosen reflection patterns are selected independent of the wireless channel. Thus, these methods don’t exploit the crucial beamforming capability of the RIS to overcome the ever-present and large RIS path losses \cite{b11}, which is included in \cite{b13} wherein the signal-to-noise ratio (SNR) needs to be extremely high, but absent in \cite{b12}.


To the best of our knowledge, this is the first work that studies RIS encoded DPolSK modulation. The main contributions of this letter are that we first show how to jointly control the SoP and beamform the scattered waves from a RIS. Then we employ this functionality by introducing a novel RIS aided information transmission approach which performs DPolSK. Finally, we validate the performance of the proposed scheme by comparing it with a benchmark scheme that performs CPolSK.

\begin{figure}[t!]
	\centering
	\includegraphics[]{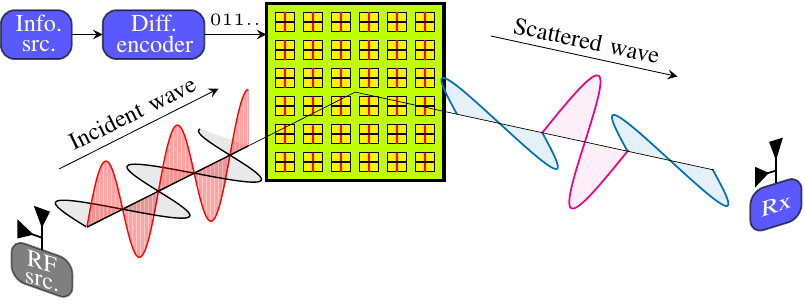}
	\caption{RIS encoded DPolSK for information transmission.} 
	\label{fig:1}
\end{figure}

\section{System Model}
\begin{figure}[t!]
	\centering
	\includegraphics[]{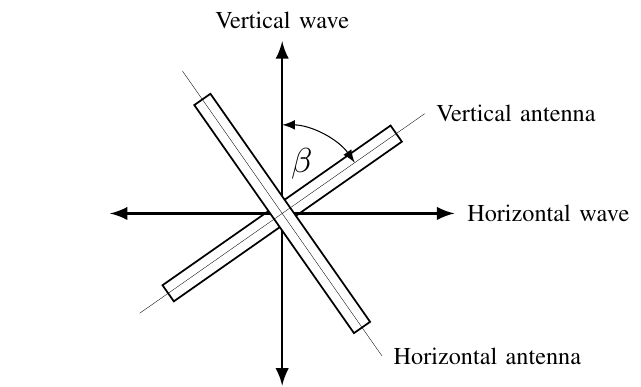}
	\caption[short text]{SoP rotation of the propagating wave at the DP receiver.} 
	\label{fig:2}
\end{figure}

A RIS with $M$ DP units transmits information to a DP receiver on impinging DP RF waves emitted from an external source, as illustrated in Fig. \ref{fig:1}. Generally, any SoP can be decomposed into any arbitrary orthogonal SoPs pair. Thus, without loss of generality, we assume that the source, receiver, and RIS units are constructed from vertical and horizontal SoPs. Furthermore, we consider only the scattered channel through the RIS; thus the direct link from the source to the receiver is neglected due to unfavourable propagation conditions such as blockage by an obstacle. Moreover, both the source and receiver are assumed to be in the far-field of the RIS. Accordingly, the received signal defined as $\mathbf{y}=[y_{_{\textnormal{V}}},y_{_{\textnormal{H}}}]^{\textnormal{T}} $, where $ y_{_{\textnormal{V}}} $ and $ y_{_{\textnormal{H}}}$ indicate the signals at the vertical and horizontal antennas, respectively, becomes
\begin{equation}
\label{1}
\mathbf{y}=\mathbf{H}\mathbf{x}+\mathbf{w},
\end{equation}
where $ \mathbf{x}=\sqrt{p_{_{t}}/2} \left[1,1\right]^{\textnormal{T}} $ is the unmodulated transmitted signal by the source of power $ p_{_{t}} $, $ \mathbf{w} \in \mathbb{C}^{2{\times}1} \sim\mathcal{N_{C}}\left(0,\sigma^2 \mathbf{I}_{2}\right)$ is the additive white Gaussian noise of variance $\sigma^{2} $, and $ \mathbf{H}\in \mathbb{C}^{2{\times}2} $ is the effective channel between the source and receiver via the scattering on the RIS units which is defined as 
\begin{equation}
\label{2}
\mathbf{H}=\sum_{m=1}^{M}  \mathbf{H}_{2,m}\Phi_{m}\mathbf{H}_{1,m},
\end{equation}
where $\Phi_{m}=\mathrm{diag} [ e^{j\varphi_{_{m,\textnormal{V}}}}, e^{j\varphi_{_{m,\textnormal{H}}}} ]$ such that $ \varphi_{_{m,\textnormal{V}}} $ and $ \varphi_{_{m,\textnormal{H}}} \in[0,2\pi]  $, $  \forall\,m\in\mathcal{M}=\left\lbrace1,2,...,M\right\rbrace $ are the phase shifts of the $ m $th unit for the vertical and horizontal SoPs, respectively, while $\mathbf{H}_{1,m}$ and $ \mathbf{H}_{2,m} \in \mathbb{C}^{2{\times}2} $ are the channels between the $m$th unit to the source and receiver, respectively. In this letter, we consider a LoS environment wherein the orthogonality between an orthogonal SoP pair is preserved while propagating in the wireless channel \cite{b14}. However, because of the distinct orientations of the DP propagating wave and destination, an SoP rotation often takes place therein as illustrated in Fig. \ref{fig:2}. Furthermore, we assume that the source to RIS channel is perfectly aligned which can be fulfilled by adjusting the orientation of the static DP RF source before operation. On the contrary, we assume for the RIS to receiver channel that there is an SoP rotation due to an azimuth deflection angle denoted by $ \beta $ between the received wave and the DP receiver as indicated in Fig. \ref{fig:2}. Consequently, the RIS to receiver channel is defined as \cite{b15}
\begin{equation}
\label{3}
\mathbf{H}_{2,m}=\rho_{_{2}} e^{-j\mu_{_2,m}} \begin{bmatrix} \cos\left(\beta\right) & \sin\left(\beta\right)\\-\sin\left(\beta\right) & \cos\left(\beta\right) \end{bmatrix} \forall\,m\in\mathcal{M}, 
\end{equation}
and given there is no SoP rotation in the source to RIS channel; $\mathbf{H}_{1,m}=\rho_{_{1}} e^{-j\mu_{_1,m}} \mathbf{I}_{2}$, $\forall\,m\in\mathcal{M}$, where $ \rho_{_{1}} $ and $ \rho_{_{2}} $ are the channel gains for the source to RIS and RIS to receiver links, respectively, whereas $ \mu_{_1,m} $ and $ \mu_{_2,m} $ are the phase shifts of the path from the $ m $th RIS unit to the source and receiver, respectively. It is noteworthy that the channel gains of the source to RIS and RIS to receiver links $ \rho_{_{1}}$ and $\rho_{_{2}}$ as well as the SoP rotation angle $\beta$  are assumed constants over the whole units in the RIS which is a feasible assumption for the far-field scenario. Thus, by the substitution of $ \mathbf{H}_{1,m} $ and $ \mathbf{H}_{2,m} $ in \eqref{2}, the received signal simplifies to
\begin{equation}
\label{4}
\mathbf{y}=\mathbf{A}\mathbf{u}+\mathbf{w},
\end{equation}
where $ \mathbf{A} $ is the SoP rotation matrix of the receiver as
\begin{equation}
\label{5}
\mathbf{A}=\begin{bmatrix} \cos\left(\beta\right) & \sin\left(\beta\right)\\-\sin\left(\beta\right) & \cos\left(\beta\right) \end{bmatrix},
\end{equation}
and $ \mathbf{u} $ accounts for the effective scattered wave from the RIS, 
\begin{equation}
\label{6}
\mathbf{u}=\sqrt{\frac{\eta^{2}p_{_{t}}}{2}}\sum\limits_{m=1}^{M}  e^{j\left( \varphi_{_{m,\textnormal{H}}}-\psi_{_{m}}\right)} \begin{bmatrix}  e^{j\Delta\varphi_{_{m}}} \\ 1 \end{bmatrix},
\end{equation}
where $\Delta\varphi_{_{m}}=\varphi_{_{m,\textnormal{V}}}-\varphi_{_{m,\textnormal{H}}}$ is the phase difference between the induced phase shifts for the vertical and horizontal SoPs by the $ m $th unit, $\eta=\rho_{_{1}}\rho_{_{2}}$ is the effective channel gain of the link created by the surface, and $\psi_{_{m}}=\mu_{_1,m}+\mu_{_2,m}$ is the effective phase shift of the path through the $ m $th unit in the surface. In this letter, for the purpose of showing the upper bound performance of the proposed method, we assume that the effective phase shifts of the paths scattered via the RIS units are known at the RIS. However, in practice, the estimation process using training signals becomes essential which is feasible with limited overhead in the LoS environment. Accordingly, the RIS is tuned to coherently combine all the scattered waves at the receiver to maximize the received signal power. Thus, from \eqref{6} the induced phase shift for the horizontal SoP and the phase difference between the induced phase shifts for vertical and horizontal SoPs become   
\begin{equation}
\label{7}
\varphi_{_{m,\textnormal{H}}}=\psi_{_{m}}\quad \text{and} \quad \Delta\varphi_{_{m}}=\Delta\varphi \quad  \forall\,m\in\mathcal{M}.
\end{equation}

Accordingly, the effective scattered wave from the surface in \eqref{6} becomes
\begin{equation}
\label{8}
\mathbf{u}=\dfrac{\alpha}{\sqrt{2}} \begin{bmatrix} e^{j\Delta\varphi} \\ 1 \end{bmatrix},
\end{equation}
where $\alpha=\sqrt{M^{2}\eta^{2}p_{_{t}}} $ and the average received SNR can be calculated as 
\begin{equation}
\label{9}
\gamma=\frac{\alpha^{2}}{2\sigma^{2}}.
\end{equation}

\begin{figure}[t!]
	\centering
	\includegraphics[]{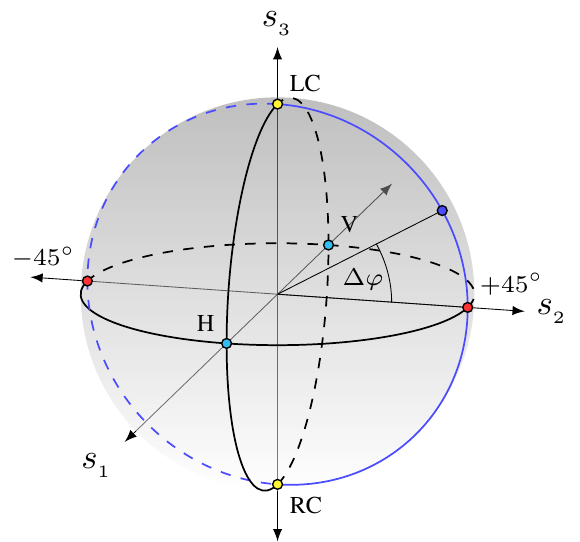}
	\caption[short text]{Poincaré sphere where the choice of $ \Delta\varphi $ controls the SoP of the effective scattered wave from RIS along the blue circle in the $s_{_{2}}$-$s_{_{3}}$ plane.} 
	\label{fig:3}
\end{figure}

In general, the amplitude ratio and the phase difference between the complex vertical and horizontal components are the two main parameters to describe any SoP \cite{b7}. Consequently, it is clear from \eqref{8} that the proper tuning of $ \Delta\varphi $ controls the SoP of the effective scattered wave from the RIS. Traditionally, the Stokes vector \cite{b16} is a fundamental tool that offers an important graphical description of the SoP. Generally, the Stokes vector of a DP signal denoted by $ \mathbf{e}=[e_{_{\textnormal{V}}},e_{_{\textnormal{H}}}]^{\textnormal{T}}$ is defined as\cite{b7}
\begin{equation}
\label{10} 
\mathbf{s}_{\mathbf{e}}=\begin{bmatrix} s_{\mathbf{e}_{_{0}}}\\ s_{\mathbf{e}_{_{1}}} \\ s_{\mathbf{e}_{_{2}}}\\s_{\mathbf{e}_{_{3}}} \end{bmatrix}=\begin{bmatrix}  \left| e_{_{\textnormal{H}}} \right|^{2} + \left| e_{_{\textnormal{V}}}  \right|^{2}\\  \left| e_{_{\textnormal{H}}} \right|^{2}- \left| e_{_{\textnormal{V}}} \right|^{2}\\ 2\Re{\{e_{_{\textnormal{H}}} e_{_{\textnormal{V}}} ^{*}}\} \\ -2\Im{\{e_{_{\textnormal{H}}} e_{_{\textnormal{V}}}^{*}}\}  \end{bmatrix},
\end{equation}
where $ \Re\{\cdot\} $, $ \Im\{\cdot\} $, and $ \{\cdot\}^{*} $ denote the real part, imaginary part, and the conjugate of their complex entries, respectively, whereas $ s_{\mathbf{e}_{_{0}}} $ represents the energy of the signal and $\mathbf{\bar{s}}_{\mathbf{e}}~=[s_{\mathbf{e}_{_{1}}},s_{\mathbf{e}_{_{2}}},s_{\mathbf{e}_{_{3}}}]^{\textnormal{T}} $ denotes the Stokes sub-vector which accounts for the SoP of the waves and allows its Cartesian mapping into the three-dimensional Poincaré space \cite{b7}. The Poincaré space offers an important geometrical interpretation of the SoP. In Fig. \ref{fig:3}, the Poincaré sphere is shown where every possible SoP can be represented by a single point on the sphere. Any two polarization states are orthogonal to each other if the line connecting them is passing through the origin. Left and right hand circular SoPs are represented by the north and south poles of the sphere on the $s_{_{3}}$-axis, respectively. In addition, all possible linear polarization states fall inside the equatorial plane where the vertical and horizontal SoPs are indicated on $ s_{_{1}} $-axis, while slant $45^{\circ}$ and slant $-45^{\circ}$ SoPs are indicated on $ s_{_{2}} $-axis.

\section{RIS Encoded Differential Polarization Shift Keying}
In this section, we highlight the capability of the RIS for controlling the SoP of the scattered waves on it. Then, we employ this function to introduce a novel RIS aided information transmission approach. In particular, the information is encoded in the SoP of the effective scattered wave from the RIS by performing a DPolSK modulation scheme. The Stokes vector of the effective scattered wave from the surface defined in \eqref{8} by following \eqref{10} becomes   
\begin{equation}
\label{11}
\mathbf{s}_{\mathbf{u}}=\begin{bmatrix} s_{\mathbf{u}_{_{0}}}\\ s_{\mathbf{u}_{_{1}}} \\ s_{\mathbf{u}_{_{2}}}\\s_{\mathbf{u}_{_{3}}}\end{bmatrix}=\alpha^{2}\begin{bmatrix} 1 \\0  \\\cos\left(\Delta\varphi\right)\\\sin\left(\Delta\varphi \right) \end{bmatrix},
\end{equation}
where it is obvious that $ \Delta\varphi  $ is the tuning parameter for the SoP of the effective scattered wave. In particular, the choice of $ \Delta\varphi\in[0,2\pi]  $ allows the manipulation of the SoP along the blue circle in the $s_{_{2}}{\text-}s_{_{3}}$ plane indicated in Fig. \ref{fig:3}. For instance; $ \Delta\varphi=0$ and $ \Delta\varphi=\pi$ create an orthogonal SoPs pair of slant~$45^{\circ}$ and slant $-45^{\circ}$, respectively. The exploitation of the RIS control over the SoP function for modulation has been limited to performing CPolSK \cite{b10}, wherein, due to the fluctuation of the SoP in the wireless channel, the estimation and correction processes of the SoP rotation become essential at the receiver. However, we introduce a novel RIS encoded DPolSK to allow non-coherent demodulation which is immune to the SoP fluctuations in the wireless channel. Specifically, in RIS encoded DPolSK, the receiver demodulates the received signal directly without the estimation and correction of the SoP rotation angle. 

In RIS encoded DPolSK, the SoPs of the effective scattered waves over two successive scattering slots are exploited to encode a single data bit. In particular, the relation between the SoP of the current effective scattered wave and that of the previous wave is used to bear the data bit. Therefore, non-coherent demodulation becomes feasible considering that the wireless channel is static during the two successive scattering slots. Furthermore, the DPolSK modulation scheme can be implemented simply using two steps. The first step is to differentially encode the data bits. While the second step is to alternate the SoP of the effective scattered waves between two possible orthogonal SoPs as a function of the differential encoded bits as illustrated in Fig. \ref{fig:1}.

Generally, the differential encoder starts with an initial encoded bit denoted by $ d_{k-1} $ at the scattering slot index $ k-1 $. Then, the information data bit denoted by $ b_{k} $ for the $ k $th scattering slot is differentially encoded as
\begin{equation}
\label{12}
d_{k}=b_{k} \mathbin{\oplus} d_{k-1},
\end{equation}
where $ \mathbf{\oplus} $ denoted the XOR Boolean operator. After that, the SoP of the effective scattered wave is chosen as a function of $ d_{k} $. In this letter, without loss of generality, we select the slant $45^{\circ}$ and slant$-45^{\circ}$ as our orthogonal SoPs pair for data encoding. However, any possible orthogonal SoPs pair will be equivalent. Consequently, the induced phase shift for the vertical SoP at the $ k $th scattering slot denoted by $ \varphi^{k}_{_{m,\textnormal{V}}} $ is tuned to switch the SoP of the effective scattering wave between slant $45^{\circ}$ and slant $-45^{\circ}$ according to the differential encoded bit as
\begin{equation}
\label{13}
\varphi^{k}_{_{m,\textnormal{V}}}=\psi_{_{m}}+\left(1-d_{k}\right)\pi \quad\;\forall\,m\in\mathcal{M}, 
\end{equation}
where \eqref{13} results in an effective scattered wave of slant $45^{\circ}$ SoP in the case of  $ d_{k}=1 $ and slant $-45^{\circ}$ SoP in the case of $ d_{k}=0 $. Furthermore, it is noteworthy that in the case of $ b_{k}=1 $, this operation switches the SoP of the $ k $th scattering slot to the second possible orthogonal SoP relative to that in the previous scattering slot. On the contrary, in the case of $ b_{k}=0 $, this operation preserves the SoP of the $ k $th scattering slot same as that of the previous scattering slot. Thus, this operation encodes the data bits differentially in the SoP of the effective scattered wave from the RIS.

Now the detection process can be done simply by observing the change in the SoPs over two successive received signals. Therefore, given that the wireless channel is static during two consecutive slots, the detection process becomes immune to the fluctuations of the wireless channel on the SoP. Furthermore, it was shown that the optimum maximum likelihood detector in the LoS environment simplifies to the minimum distance detector in the Poincaré space \cite{b17}. Thus, the change in the SoPs can be tracked using the Stock sub-vectors of two successive received signals, and the demodulation process for the $ k $th information bit becomes \cite{b18} 
\begin{equation}
\label{14}
\hat{b}_{k}=
\begin{cases}
0 & \text{ if } \mathbf{\bar{s}}_{\mathbf{y}_{_{k}}}^{\textnormal{T}}\mathbf{\bar{s}}_{\mathbf{y}_{_{k-1}}} \geq 0\\
1 & \text{ if } \mathbf{\bar{s}}_{\mathbf{y}_{_{k}}}^{\textnormal{T}}\mathbf{\bar{s}}_{\mathbf{y}_{_{k-1}}} < 0
\end{cases},
\end{equation}
where $\mathbf{\bar{s}}_{\mathbf{y}_{_{k}}}$ and $ \mathbf{\bar{s}}_{\mathbf{y}_{_{k-1}}}$ are the Stoke sub-vectors for the two successive received signals $ \mathbf{y}_{_{k}} $ and  $ \mathbf{y}_{_{k-1}} $, respectively. It is noteworthy that the detection process needs two consecutive received signals. Therefore, the performance is affected by two independent noise samples which result in performance degradation in comparison to the CPolSK. However, the detection can be done directly based on the received signals without the SoP rotation angle estimation and compensation processes in the receiver. It was shown in \cite{b18} that the theoretical bit-error-rate (BER) of the DPolSK is 
\begin{multline}
\label{15}
\textnormal{BER}=\frac{1}{2\pi} \int\displaylimits_0^{2\pi} \int\displaylimits_0^{\infty} f_{\eta}(\eta)\left[1- F_{\vartheta}\left(\acot\left[ \frac{\cos(\delta)}{\eta}\right]  \right)  \right]  \mathrm{d\eta} \mathrm{d\delta} \\+\frac{1}{2\pi} \int\displaylimits_0^{2\pi} \int\displaylimits_{-\infty}^0 f_{\eta}(\eta) F_{\vartheta}\left(\acot\left[ \frac{\cos(\delta)}{\eta}\right]  \right)   \mathrm{d\eta} \mathrm{d\delta} ,
\end{multline} 
where the probability density function $ f_{\eta}(\eta) $ and the cumulative distribution function $ F_{\vartheta}(\vartheta) $ are defined as \cite{b18}
\begin{multline}
\label{16}
 f_{\eta}(\eta)=\frac{1}{2}\left( \frac{1}{1+\eta^{2}}\right)^{\frac{3}{2}} e^{-\gamma\left(1-\frac{\eta}{\sqrt{1+\eta^{2}}} \right) }\\\cdot\left[1+\gamma\left(1+\frac{\eta}{\sqrt{1+\eta^{2}}} \right)  \right] \;\eta \in[-\infty,\infty]   ,
\end{multline}
\begin{equation}
\label{17}
F_{\vartheta}(\vartheta)=1-\frac{1}{2}e^{-\gamma\left(1-\cos\vartheta\right) } \left(1+\cos\vartheta\right)  \;\vartheta \in[0,\pi],
\end{equation}   
where $ \gamma $ is the received SNR defined in \eqref{9}.

\section{Performance Evaluation}
In this section, we evaluate the performance of the proposed RIS encoded DPolSK modulation scheme. The simulation parameters are presented in Table \ref{table:1}. Given the RIS far-field scenario, we depend on the plate-scattering path-loss model and the radiation pattern of the reflecting unit presented in \cite{b11}. Accordingly, the effective channel gain of the link created by the surface is
\begin{equation}
\label{18}
\eta=\left( \dfrac{\Delta\sqrt{G_{t}G_{r}}}{4\pi r_{_{1}}r_{_{2}}}\right) \left[ \cos(\zeta_{_{1}})\cos(\zeta_{_{2}})\right] ^{q_{_{o}}} ,
\end{equation}
where $q_{_{o}}=0.285 $ for the square shape reflecting unit of side length equals to half-wavelength \cite{b11}, while $ G_{t} $ and $ G_{r} $ are the gains of the source and receiver antennas respectively, whereas $ \Delta $, $r_{_{1}}$ and $ r_{_{2}}$ are the reflecting unit's physical area, the distance from the source to RIS, and the distance from the RIS to receiver, respectively. Furthermore, $\zeta_{_{1}}$ and $ \zeta_{_{2}}$ are the angles between the normal of the RIS to the incident and scattered waves, respectively. It is noteworthy that the received SNR in \eqref{9} is proportional to $ A_{\textnormal{RIS}}^{2}  $ where $ A_{\textnormal{RIS}}=M\Delta $ is the RIS' physical area. Therefore, for the sake of neutralizing the carrier frequency effect on the performance, we use in these simulations the RIS' physical area instead of its number of units.

\begin{table}[t]
	\label{table}
	\centering
	\caption{Simulation Parameters}
	\begin{tabular}{r|r}
		\textbf{Parameter}&\textbf{Value}  \\ \hline
		Gain of transmit and receive antennas &$ 3 $ dBi \\ 
		Carrier frequency& $ 3 $ GHz \\ 
		Transmission power ($p_{_{t}}$)&8 dBm \\ 
		Noise power ($\sigma^2$)&-96 dBm \\ 
		Reflecting unit physical area ($\Delta$)&$ \lambda_{c}/2 \times \lambda_{c}/2 $ \\ 
		Location of RF source&$ \left[50, 0, 0\right]m$  \\ 
		Location of Receiver &$ \left[50, 100, 0\right]m$  \\ 
		Location of RIS &$ \left[0, 50, 0\right]m$  
		\label{table:1}
	\end{tabular}
\end{table}

Moreover, given the far-field scenario, the phase shifts of the paths from the RIS units to the source and receiver are computed using the plane wave propagation model as \cite{b19}      
\begin{equation}
\label{19}
\mu_{_{l,m}}=\mathbf{g}_{m}^\mathrm{T}\mathbf{q}_{_{_{l}}} \quad \forall\,m\in\mathcal{M}, 
\end{equation}
where $ l\in\left\lbrace 1,2\right\rbrace $, and
\begin{gather}
\label{20}
\mathbf{q}_{_{_{l}}}=\frac{2\pi}{\lambda_{c}}\left[ 
\cos\left(\phi_{_{l}}\right)\cos\left(\theta_{_{l}}\right),\sin\left(\phi_{_{l}}\right)\cos\left(\theta_{_{l}}\right),\sin\left(\theta_{_{l}}\right) \right]^{\textnormal{T}},
\end{gather}
where $ \lambda_{c} $ is the carrier wave-length, $\mathbf{g}_{m} \in\mathbb{R}^{3{\times}1} $ is the Cartesian coordinates of the $m$th unit in the surface, and $\mathbf{q}_{_{_{l}}}\in \mathbb{R}^{3{\times}1}$ is the wave vector which accounts for the phase variations over the RIS units for the incident and reflected waves, while $\theta_{_{l}}$ and $\phi_{_{l}}$ are the elevation and azimuth angles of arrival and departure of the RIS, respectively.

\setlength{\textfloatsep}{0pt}  
\begin{figure}[t!]
	\centering
	\includegraphics[width=8.5cm]{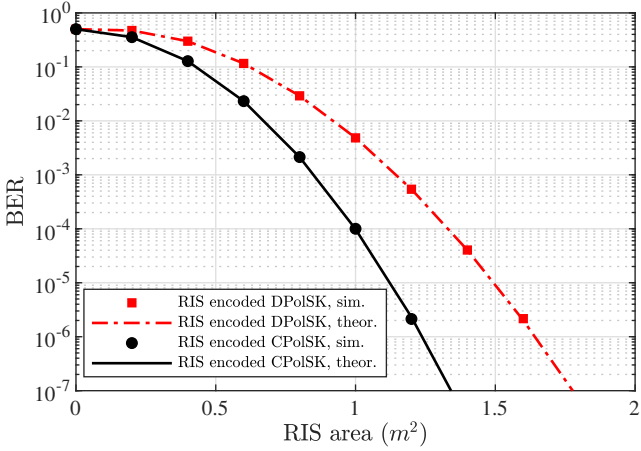}
	\caption[short text]{BER performances of RIS encoded DPolSK and CPolSK schemes are shown versus the RIS area.}
	\label{fig:4}
\end{figure}

\begin{figure}[t!]
	\centering
	\includegraphics[width=8.5cm]{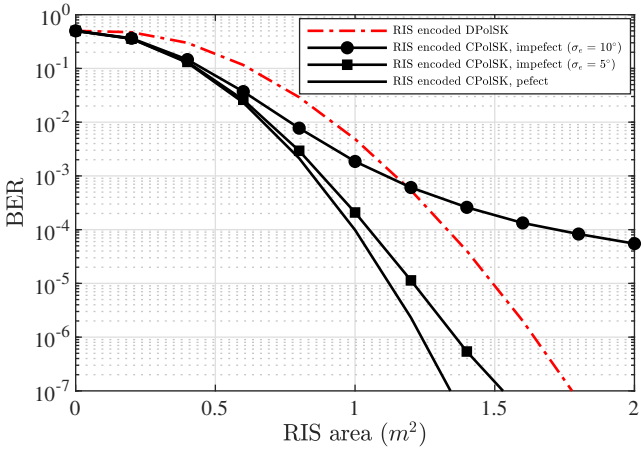}
	\caption[short text]{BER performances of RIS encoded DPolSK and CPolSK schemes are shown versus the RIS area given there are SoP rotation angle estimation errors.}
	\label{fig:5}
\end{figure}

In Fig. \ref{fig:4}, the theoretical and simulated BER performance for the proposed RIS encoded DPolSK modulation scheme are compared against the benchmark RIS encoded CPolSK scheme. In the case of CPolSK, the RIS switches the SoP of the scattered wave between two orthogonal SoPs as a function of the data bits. Then, the receiver initially corrects for the SoP rotation of the wireless channel and after that the data bit detection is processed as shown in \cite{b10}. It is shown that the theoretical BER performance of CPolSK is $ \textnormal{BER}=0.5\,e^{-\gamma} $\cite{b18}. In addition, the theoretical BER performance of the RIS encoded DPolSK in \eqref{15} is computed numerically. Although the performance of CPolSK is better than that of DPolSK, the DPolSK allows non-coherent detection without the need for the estimation and compensation processes for the SoP rotation which takes place in the wireless channel at the receiver.

In Fig. \ref{fig:5}, the BER performances are shown in the scenario where there are estimation errors in the SoP rotation angle at the receiver. We model the estimation errors as a zero-mean Gaussian distribution of standard deviation denoted by $ \sigma_{e} $ that represents the estimator accuracy. The performance of RIS encoded DPolSK is independent of the SoP fluctuations of the wireless channel as it uses non-coherent detection. However, it is obvious that the performance of RIS encoded CPolSK is sensitive to the SoP rotation angle estimation errors. Therefore, the advantage of the DPolSK arises over CPolSK in the presence of SoP rotation angle estimation errors at the receiver.

\section{Conclusion}
We proposed a novel RIS encoded information transmission scheme for LoS environment by utilizing the RIS capability to control the SoP of the scattered wave to perform DPolSK modulation scheme. In particular, the SoPs of the scattered waves from the RIS are alternated differentially over two successive scattering slots to encode the information data bit. With a performance penalty compared to RIS encoded CPolSK, the proposed scheme allows non-coherent detection without the need for estimation and compensation of the SoP rotation at the receiver.

\bibliographystyle{ieeetran}
\bibliography{references}

\end{document}